\documentclass[fleqn,12pt,twoside]{article}
\usepackage[headings]{espcrc1}
\readRCS
$Id: espcrc1.tex,v 1.2 2004/02/24 11:22:11 spepping Exp $
\usepackage{epsf}

\title{Statistical analysis of structural transitions in small systems}

\author{Michael Bachmann\address{
Institut f\"ur Festk\"orperforschung, Theorie II, Forschungszentrum J\"ulich,\\
D-52425 J\"ulich, Germany\\
E-mail: m.bachmann@fz-juelich.de
}}

\runtitle{Statistical analysis of structural transitions in small systems}
\runauthor{M.\ Bachmann}

\begin{document}
\maketitle
\begin{abstract}
We discuss general thermodynamic properties of molecular structure formation processes
like protein folding by means 
of simplified, coarse-grained models. 
The conformational
transitions accompanying these processes exhibit similarities to thermodynamic phase transitions,
but also significant differences as the systems that we investigate here are very small. 
The usefulness of a microcanonical statistical analysis of these transitions
in comparison with a canonical interpretation is emphasized. The results 
are obtained by employing sophisticated 
generalized-ensemble Markov-chain Monte Carlo methodologies.
\end{abstract}

\section{Polymers and proteins}
\label{bj:sec1}

In heterogeneous many-particle systems, where typically different energy scales, associated with the
inter-particle interactions, compete with each other, the formation of stable composites
is a complex process. This is even more relevant if the many-particle system exhibits 
constraints such as covalent chemical bonds restricting the mobility of atomistic or monomeric 
subunits. Furthermore, the flexibility of 
the molecular system to adapt to environmental conditions (solvent, temperature, pressure, etc.)
is reduced. Non-covalent interactions among such subunits which are due, for example, to
dipole--dipole interactions (van der Waals ``bonds'', hydrogen bonds), cause additional 
limitations~\cite{rubinstein1}. 

Concentrating on linear, linelike molecules in solvent, 
two major classes are discriminated: homopolymers and
heteropolymers. Homopolymers are chains of repeated identical subunits, i.e., chemical groups
(``monomers''); a typical example is
polyethylene -[CH$_2$]-. The ``two ends'' of the backbone of the characteristic monomers 
(in the case
of polyethylene this is the methylene group CH$_2$) are connected by covalent bonds 
(e.g., C-C) with the previous and the next 
group, respectively. 

In general, depending on symmetry and stability, covalent bonds
strongly differ in their response to excitations, which, for example, can be
induced by external fields or thermal 
fluctuations. If, under present environmental conditions and external fields, these bonds are hardly 
excited to perform fluctuations, then they are rather ``stiff'', i.e., the contour length of
the chain is widely conserved. Thus, depending
on the backbone stiffness, longitudinal degrees of freedom are frozen and thus result in constraints
of the tangential monomeric mobility. In other cases, polymers can be elastic and due to the 
longitudinal flexibility, the contour length fluctuates upon excitation. 

Basic backbone geometries that can be associated with such linelike objects are helices, strands,
rodlike, and ``random'' wormlike conformations. Helices and sheets belong to the class
of secondary structures
and are stabilized by hydrogen bonds formed by dipole--dipole interactions of polar groups 
(e.g., -CO, -NH, -OH) in the polymer backbone. However, it shcould be noted that
the formation of hydrogen bonds is not
a necessary condition for the onset of secondary structures which are rather 
elementary intrinsic geometries of such linear monomeric chains. This is particularly visible,
if the monomers contain side chains whihc can be modeled as tubelike polymers~\cite{banavar1,vbj1}.

If the monomers have a larger chemical complexity,
e.g., by the existence of side chains (e.g., polar groups, aromatic rings, etc.) protruding from the backbone, 
repulsive volume exclusion
effects and attractive interactions among side chains (caused by
electrostatic, van der Waals, or hydrophobic forces), higher-order (``tertiary''),
i.e., crystalline or amorphous, glassy structures can form.
These interactions entail a further, not necessarily uniform, limitation
of the individual monomeric mobility. Typically, the energy of such non-covalent bonds
among side chains 
of different monomers is typically sufficiently small and non-covalent bonds can break by 
comparatively weak changes of environmental conditions.
Consequently, different phases of polymers are discriminated which depend on
temperature, polymer concentration, solvent quality, pressure, presence of external fields, etc.

However, despite the huge number of natural and technologically important synthesized polymers 
with different chemical composition, a rough classification of linear polymers into
groups of flexible, semiflexible, and stiff polymers is possible. Thus, only a few 
system parameters, based on the fundamental length and energy scales associated with
covalent bonds and non-covalent interactions, are necessary to describe the generic phase behavior of 
classes of polymers~\cite{rubinstein1}. Conformational transitions of single homopolymers can be understood as 
thermodynamic phase transitions if the degree of polymerization, i.e., the length of the 
chain or the number of monomers $N$, is huge~\cite{deGennes1,flory1}. 
The general structural behavior of polymers
is typically described by geometric quantities such as the end-to-end distance $R_{\rm ee}$ and 
the radius of gyration $R_{\rm gyr}$, the latter being a measure for the compactness of
polymers. For very long chains, i.e., in the limit $N\to\infty$, the ensemble average
has a power-law form, 
$\langle R_{\rm ee,gyr}^2\rangle\sim N^{2\nu}$, where $\nu$ is a universal exponent which is a
characteristic constant in the structural phase the polymer resides in. For a flexible
interacting polymer in good solvent (or temperatures $T$ above the $\Theta$ point $T_\Theta$), 
for example, $\nu\approx 3/5$. This corresponds to the 
behavior of self-avoiding walks (random coils with pure steric volume exclusion).   
At the $\Theta$ point that separates
good from poor solvent conditions, or alternatively, the random-coil phase from the globular phase of
compact conformations at $T_\Theta$, $\nu=1/2$. The polymer behaves like a random walk~--
repulsive volume exclusion and attractive monomer--monomer interaction cancel each other.
In the globular and crystalline phase ($T<T_\Theta$), very compact (or spherical) conformations
dominate and thus $\nu\approx 1/3$.

However, it should be noted that the interest in polymers of rather small length $N\ll \infty$ has 
increased in the past years with the onset of nowadays available high-resolution experimental techniques
and the demand
for nanofabrication of molecular applications. The investigation of the
finite-length behavior of single polymers sheds some new light on the conformational transitions
of polymers.
Nucleation effects are governed by the competition of likewise surface and volume effects as well
as entropic ambiguity and energetic significance of polymer conformations in a certain macrostate
(``pseudophase'') and close to transitions between different macrostates (``pseudophase transitions'').
The thermodynamic limit is not of relevance anymore and thus also is not the asymptotic critical
behavior. The understanding of subphases and transitions in-between becomes crucial, but this
is challenging because of the high specificity of small polymers. Thus, a generic classification of 
conformational subphases is
an intricate problem as details of monomeric properties~-- even down to the atomistic level --
can essentially influence the overall pseudophase behavior. The description of structural 
behavior by means of theoretical models is, therefore, difficult and one has to be aware that
results of studies of the same polymer can differ from model to model. Thus, model
studies of finite-length polymers are of highly qualitative nature and results have to be
interpreted as ``possible scenarios'' rather than definite answers~\cite{bjX1}.

This is particularly relevant for a special class of polymers, the heteropolymers. 
The most prominent representatives are the proteins which are the workhorses 
in all biological cell systems. Proteins are synthesized by the ribosomes in the cell, 
where the genetic code in the DNA 
is translated into a sequence of amino acids. The folding of a synthesized protein into
its three-dimensional structure is typically a spontaneous process that takes place
in an aqueous environment. 
In a complex biological system, the large variety of processes which are necessary to 
keep an organism alive requires an ensemble of different functional proteins. In the 
human body, for example, about 100\,000 different proteins fulfill specific functions.
However, this number is extremely small, compared to the huge number of possible 
amino acid sequences ($=20^N$, where $N$ is the number of amino acids, typically in the range
100 to 3000).
The reason is that bioproteins have to obey very specific demands. Most important are
stability, uniqueness, and functionality.

\section{Folding transitions of proteins}
\begin{figure}[t]
\centerline{\epsfxsize=5cm \epsfbox{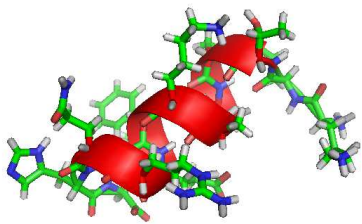}}
\caption{\label{fig:cpep} 
\small The C-peptide of ribonuclease A consists of 13 amino acids and is a typical $\alpha$-helix
former. 
}
\end{figure}

The structural changes a protein experiences in the folding process are of different 
nature. There are rather local, but nonetheless cooperative arrangements of monomeric subunits
in helical or sheetlike segments. These are \emph{secondary structures} (the amino acid sequence is
the \emph{primary structure}). The formation of secondary structures is a conformational transition;
a very prominent one is the helix-coil transition: In a collective effect of orientational
ordering, an unfolded segment of adjacent amino acids transforms into a helical substructure.
Molecular helix-coil transitions are typically accompanied by hydrogen-bond formation. Hydrogen
bonds stabilize the symmetry of secondary structures. A famous example is the Watson-Crick
$\alpha$-helix with 3.6 amino acids per winding (see Fig.~\ref{fig:cpep}). 

\begin{figure}[t]
\centerline{\epsfxsize=10cm \epsfbox{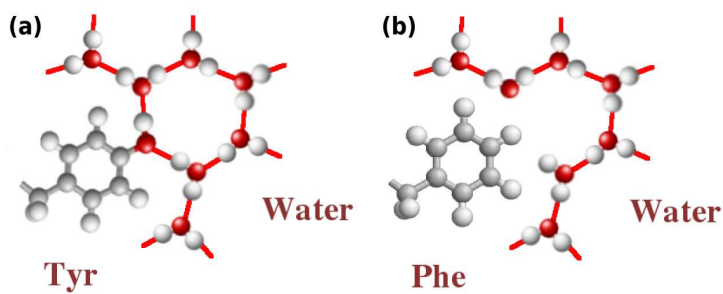}}
\caption{\label{fig:hpf} 
\small (a) Tyrosine (Tyr) is an amino acid with an OH group in the side chain. Thus, it is hydrophilic
as the OH dipole can form a hydrogen bond with a polar solvent molecule (water). (b)
Phenylalanine (Phe), on the other hand, is a typical example for a hydrophobic amino acid.
The CH$_2$-C$_6$H$_5$ side chain does not contain a polar group. Phe in the surface-accessible
protein shell would disturb the hydrogen-bond network of the solvent which is energetically disfavored.
}
\end{figure}

In proteins, the size of individual secondary-structure segments is typically
rather small. The reason is
that proteins are ``interacting polymers'', i.e., the amino acids strongly interact with
each other and form a globular or \emph{tertiary} shape. This is due to the fact that amino acids 
possess a uniform backbone and differ only in their side chains. Adjacent amino acid 
backbones are connected via the 
peptide bond and electric dipoles formed by backbone atoms are typically involved in
hydrogen-bond formation. Backbone-backbone interaction provides the symmetry of secondary structures. 
However, the interaction between the non-bonded side chains is non-uniform and strongly 
dependent of the side chain type. Roughly, two significantly different classes of
side chains occur: hydrophilic ones that favor contact with a surrounding polar solvent like
water and hydrophobic side chains which are non-polar, thus disfavoring contact with water
molecules (for representatives of the two classes see Fig.~\ref{fig:hpf}). 
Therefore, the effective force that leads to the formation of a compact hydrophobic core 
surrounded by a screening shell of polar amino acids is called hydrophobic force. For 
spontaneously folding single-domain proteins it is the essential driving force in the tertiary
folding process.

The folded structure of a functional bioprotein is thermodynamically stable under physiological
conditions, i.e., thermal fluctuations do not lead to significant globular conformational changes.
To force tertiary unfolding requires an activation energy that is much larger than the
energy of the thermal fluctuations. This activation barrier can be drastically reduced
by the influence of other proteins, the prions. The Creutzfeld-Jakob disease is
an example for the disastrous consequences prion-mediated degeneration of proteins 
can cause in the brain. 
The folded structure and the statistical ensemble of native-like structures, which are 
morphologically identical to the native fold, form a macrostate. It represents a
conformational phase which is energy-dominated. Functionality of the native structure
is only assured if entropic effects are of little relevance.

A significant change of the environmental conditions such as temperature, pH value, or, 
following the above example, the prion concentration, can destabilize the folded phase.
Entropy becomes relevant, the entropic contribution to the free energy
starts dominating over energy. Consequently, the hydrophobic core decays. This does not necessarily
lead to a globular unfolding of the protein. A rather compact intermediate conformational
phase can be stable~\cite{sbj1}. However, further imbalancing the conditions will finally lead 
to the phase of randomly unstructured coils. The latter transition is often called
``folding/unfolding transition'', whereas the hydrophobic core formation is referred to 
as ``glassy transition'', as unresolvable competing energetic effects may result
in frustration. 
The primary structure, i.e., the sequence of different amino acids lining up in proteins,
is already sort of quenched disorder. 
\begin{figure}[t]
\centerline{\epsfxsize=12cm \epsfbox{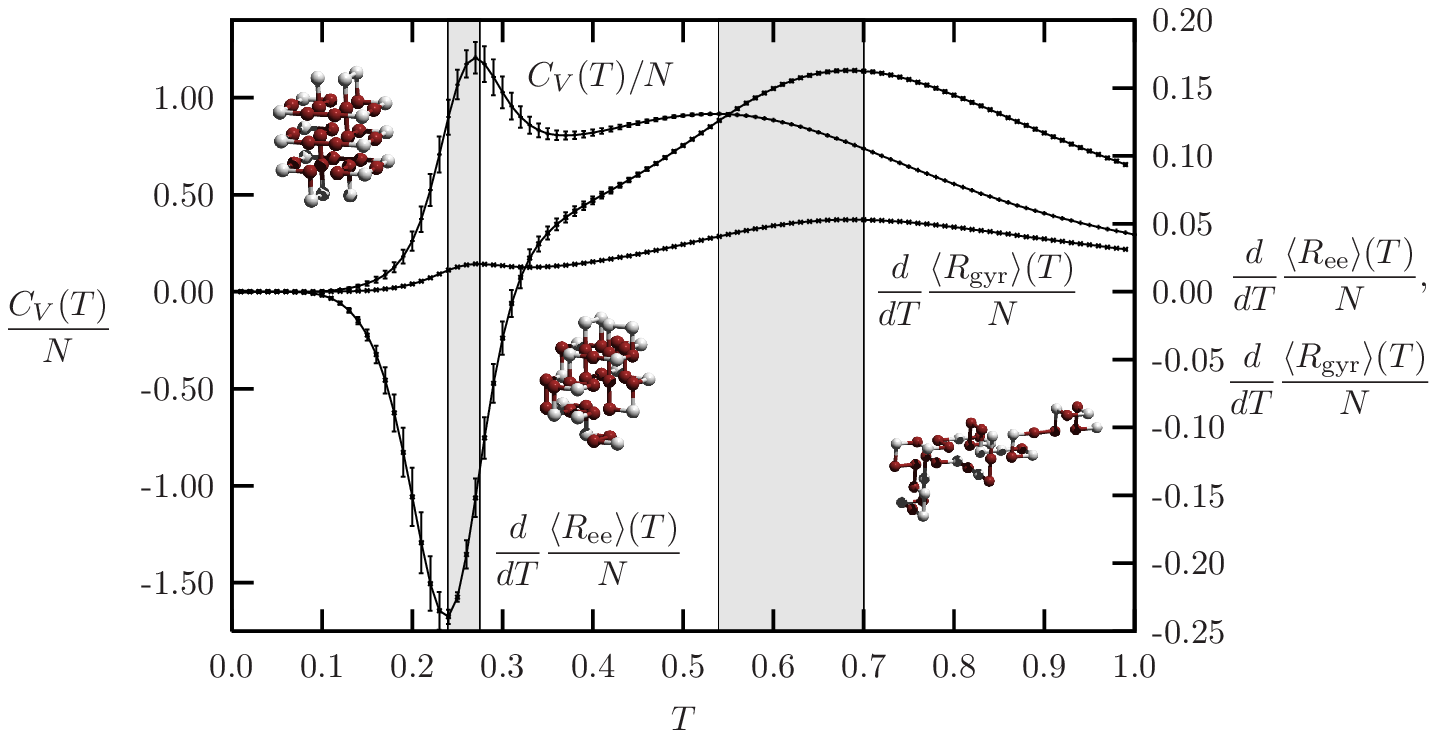}}
\caption{\label{fig:hpfold} 
\small Folding thermodynamics of a 42-residue peptide using a simple hydrophobic-polar lattice peptide
model~\cite{bj1}. Shown are the specific heat and 
fluctuations of structural quantities such as end-to-end distance and radius of gyration.
}
\end{figure}

A simple example for these transitions accompanying a lattice peptide~\cite{dill1} folding process is shown in
Fig.~\ref{fig:hpfold}, where the specific heat $C_V$ and the fluctuations of end-to-end distance
and radius of gyration, $d\langle R_{\rm ee,gyr}\rangle/dT$, are plotted as functions of temperature. 
Above the folding transition, structureless conformations dominate the ``vapor phase''.
Below the folding transition, but above the hydrophobic-core collapse, very compact
``globular'' structures form the ``liquid phase''. When, for lower temperatures, the hydrophobic-core
formation proceeds, the fluctuations get smaller. Finally, the native state with maximally
compact hydrophobic core has formed. As in this example, the native state is not necessarily
the overall most compact conformation. This is obvious from the negativity of the
gyration radius fluctuations in this transition region. Although all fluctuating quantities
clearly signalize the transitions, the peak temperatures do not perfectly coincide. 
This is a typical indication 
for the finiteness of the system. There are no transition points in protein structure-formation 
processes, but rather transition regions~\cite{bjX1,bj1}. This separates conformational transitions of
finite-length polymers (pseudophase transitions) from thermodynamic phase transitions 
being considered in the thermodynamic limit.

\section{Microcanonical vs.\ canonical interpretation}

The smallness of such systems can cause surprising side-effects in nucleation processes
which protein folding belongs to. Since the formation of the solvent-accessible hydrophilic surface
and the bulky hydrophobic core is crucial for the whole tertiary folding process, the 
competition between surface and volume effects significantly influences the thermodynamics
of nucleation~\cite{gross1,hilbert1,behringer1,jbj1,chen1}. 
For this reason, it is not obvious at all, which statistical ensemble
represents the appropriate frame for the thermodynamic analysis of folding processes.
This is even more intricate, as one may think. It is, for example, quite common to interpret
phase transitions by means of fluctuating quantities calculated within the canonical 
formalism. Transition points are characterized by divergences in the fluctuations (second-order
phase transitions) or entropy discontinuities (first-order transitions), occurring at 
unique transition temperatures. This standard analysis is based on the assumption that the
temperature is a well-defined quantity, as it seems to be an easily accessible control
parameter in experiments. This assumption is true for very large systems ($N\to\infty$) with vanishing 
surface/volume ratio in equilibrium, where surface fluctuations are irrelevant. The microcanonical
Hertz entropy $S(E)=\ln\, G(E)$ (with $k_B=1$ in our units), where $G(E)=\int_{E_{\rm min}}^E
dE' g(E')$ is the integrated density of states, is a concave function and thus the 
microcanonical temperature, defined by the mapping
${\cal F}:E\mapsto T=:T(E)=[\partial S(E)/\partial E]^{-1}$, never decreases 
with increasing energy $E$. A discrimination of the parameter ``temperature'' in the canonical ensemble
and the microcanonical (caloric) temperature $T(E)$ is not necessary, as energetic fluctuations
vanish and thus the canonical and the microcanonical ensemble are equivalent
in the thermodynamic limit.

But what if surface fluctuations are non-negligible? In this case,
the canonical temperature can be a badly defined control parameter for studies of nucleation
transitions with phase separation.\footnote{Folding or ``nucleation'' processes of proteins are 
strongly dependent on the sequence of amino acids. Thus,
folding is no generic phase transition and terms like ``nucleation'' should be used 
with some care.} This becomes apparent in the following microcanonical folding analysis 
of an exemplified off-lattice hydrophobic-polar heteropolymer with 20 monomers
and sequence H$_3$P$_2$HP$_2$HPHP$_2$HPHPHPH, as described by the AB model~\cite{still1,hsu1,baj1}. 

\begin{figure}[t]
\centerline{\epsfxsize=12cm \epsfbox{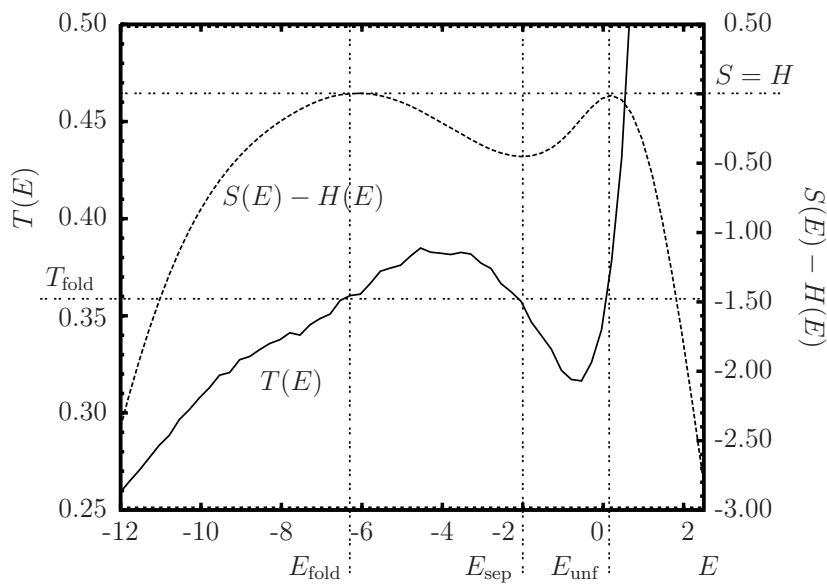}}
\caption{\label{fig:micro} 
\small Folding transition as a phase-separation process for an exemplified 20mer
in the AB model. In the transition region, the caloric temperature 
$T(E)$ of the 
protein decreases with increasing total energy. Folded and unfolded  
conformations coexist at the folding transition temperature, where
$S(E)=H(E)$, corresponding to $T_{\rm fold}\approx 0.36$. 
The energetic transition region is bounded by the
energies $E_{\rm fold}\approx -6.3$ and $E_{\rm unf}\approx 0.15$. Folding and unfolding
regions are separated at $E_{\rm sep}\approx -2.0$. 
}
\end{figure}

From multicanonical~\cite{muca1}
computer simulations, an accurate estimate of the density of states $g(E)$ can be obtained.
For this particular heteropolymer, it turns out that the entropy $S(E)$ exhibits a \emph{convex}
region, i.e., a tangent with two touching points, at $E_{\rm fold}$ and $E_{\rm unf}>E_{\rm fold}$,
can be constructed. This so-called Gibbs hull is then parametrized by 
$H(E)=S(E_{\rm fold})+E/T_{\rm fold}$, where 
$T_{\rm fold}=[\partial H(E)/\partial E]^{-1}=[(\partial S(E)/\partial E)_{E_{\rm fold},E_{\rm unf}}]^{-1}$
is the microcanonically defined folding temperature, which is here $T_{\rm fold}\approx 0.36$.
As shown in Fig.~\ref{fig:micro}, the difference $S(E)-H(E)$ has two zeros at
$E_{\rm fold}$ and $E_{\rm unf}$, and a
noticeable well in-between with the local minimum at $E_{\rm sep}$. The deviation
$\Delta S=H(E_{\rm sep})-S(E_{\rm sep})$ is called \emph{surface entropy} as the 
convexity of the entropy in this region is caused by surface effects.

\begin{figure}[t]
\centerline{\epsfxsize=12cm \epsfbox{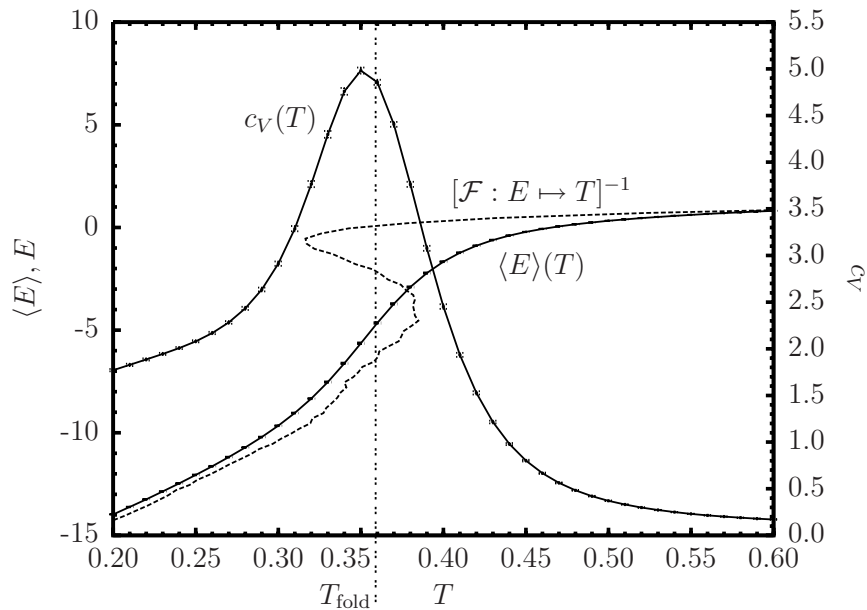}}
\caption{\label{fig:canon} 
\small Comparison of canonical and microcanonical analysis for the folding transition:
The fluctuations of energy as represented by the specific heat per monomer, $c_V(T)$,
exhibit a sharp peak near $T\approx 0.35$. The canonical mean energy $\langle E\rangle(T)$ 
crosses over from the folded to the unfolded conformations. However, the canonical
calculation averages out the overheating/undercooling branches and the backbending effect which
are clearly signaled by the microcanonical analysis of the (inverse) mapping between temperature and
energy. The microcanonically defined ``folding temperature'' $T_{\rm fold}$ is close to the peak temperature 
of the specific heat.
}
\end{figure}

However, the most striking feature in Fig.~\ref{fig:micro} is the qualitative change of the 
microcanonical temperature $T(E)$ in the transition region: Approaching from small energies
(folded phase), the curve passes the folding temperature $T_{\rm fold}$ and follows the overheating branch. 
It then \emph{decreases} with \emph{increasing energy} (passing again $T_{\rm fold})$ before
following the undercooling branch, crossing $T_{\rm fold}$ for the third time. In the unfolded
phase, temperature and energy increase as expected. The unusual backbending of the caloric temperature
curve within the transition region is not an artifact of the theory. It is a physical effect and
has been confirmed in sodium cluster formation experiments~\cite{schmidt1}, where a similar behavior
was observed.\footnote{It is sometimes argued that 
proteins fold in solvent, where the solvent serves as heat bath. This would provide a fixed canonical 
temperature such that the canonical interpretation is sufficient to understand the transition.
However, the solvent-protein interaction is actually implicitly contained in the heteropolymer
model and, nonetheless, the microcanonical analysis reveals this effect which is simply 
``lost'' by integrating out the energetic fluctuations in the canonical ensemble (see Fig.~\ref{fig:canon}).} 

In Fig.~\ref{fig:canon}, results from the canonical calculations (mean energy $\langle E\rangle$
and specific heat per monomer $c_V$) are shown as functions of the temperature. The specific heat
exhibits a clear peak near $T=0.35$ which is close to the folding temperature $T_{\rm fold}$, as
defined before in the microcanonical analysis. The loss of information by the canonical averaging
process is apparent by comparing $\langle E\rangle$ and the inverse, non-unique mapping 
${\cal F}^{-1}$ of microcanonical temperature and energy. The temperature decrease 
in the transition region from the folded to the unfolded structures is unseen in 
the plot of $\langle E\rangle$. 

Eventually, as we had already mentioned when discussing the results shown in Fig.~\ref{fig:hpfold}, 
there is also no unique canonical 
folding temperature signaled by
peaks of fluctuating quantities; there are rather transition regions. 
Therefore, for small systems, the definition of transitions based on the 
canonical temperature is indeed little useful. Consequently, the microcanonical
analysis is at least a powerful alternative for discussing conformational transitions
of ``small'' systems. This also applies to molecular aggregation~\cite{jbj1} and 
adsorption transitions~\cite{mbj1}, where
similar phenomena can occur.

\section{Conclusions}
As we have exemplarily shown for the folding of proteins, 
conformational transitions of molecular systems can be
well-described by analysis techniques known from statistical physics. However, the 
interpretation of these cooperative effects as thermodynamic phase transitions has
noticeable limitations. This not only regards the impossibility to precisely identify
definite transition points. It is even more fundamental to ask the question which of
the typically used statistical ensembles provides the most comprehensive interpretation
of finite-system structure formation processes. We have shown that the microcanonical
analysis of folding thermodynamics is particularly advantageous, as, e.g., the  
remarkable temperature backbending effect is averaged out in the canonical ensembles, 
where the temperature is considered as an external control parameter which seems
to be questionable for small systems~\cite{gross1,hilbert1,behringer1,jbj1,chen1}.

I thank Wolfhard Janke for quite successful long-term collaboration. This work is partially supported by
the DFG (German Science
Foundation) under Grant Nos.\ JA 483/24-1/2/3. Support
by a supercomputer time grant of the John von Neumann
Institute for Computing (NIC), Forschungszentrum J\"ulich,
is acknowledged.
\end{document}